# Laboratory Demonstration of an Active Optics System for High-Resolution Deployable CubeSat

Noah Schwartz [(1)], David Pearson [(1)], Stephen Todd [(1)], Maria Milanova [(1)],
William Brzozowski [(1)], Andy Vick [(2)], David Lunney [(1)], Donald MacLeod [(1)],
Steve Greenland [(1)], Jean-François Sauvage [(3)], Benjamin Gore [(1, 4)]

[(1)] UK Astronomy Technology Centre, Blackford Hill, Edinburgh EH9 3HJ, United Kingdom,
+44 (0)131 668 8256, noah.schwartz@stfc.ac.uk.
[(2)] RAL Space, Fermi Ave, Harwell, Didcot OX11 0QX, United Kingdom.
[(3)] ONERA, 29 avenue de la Division Leclerc, 92322 Châtillon, France.
[(4)] Heriot-Watt University, Edinburgh Campus, Edinburgh EH14 4AS, United Kingdom.

## ABSTRACT

In this paper we present HighRes: a laboratory demonstration of a 3U CubeSat with a deployable primary mirror that has the potential of achieving high-resolution imaging for Earth Observation. The system is based on a Cassegrain telescope with a segmented primary mirror composed of 4 petals that form an effective aperture of 300 mm. The design provides diffraction limited performance over the entire field-of-view and allows for a panchromatic ground-sampling distance of less than 1 m at an altitude of 350 km. The alignment and co-phasing of the mirror segments is performed by focal plane sharpening and is validated through rigorous numerical simulations. The opto-mechanical design of the prototype and its laboratory demonstration are described and measurements from the on-board metrology sensors are presented. This data verifies that the performance of the mirror deployment and manipulation systems is sufficient for co-phasing. In addition, it is shown that the mirrors can be driven to any target position with an accuracy of 25 nm using closed-loop feedback between the mirror motors and the on-board metrology.

## 1   INTRODUCTION

Since the initial proposition by Twiggs et al. [1] and first flights in 2003 [2], the capability of CubeSat systems has been steadily increasing. Three drivers dominate the image quality available from a satellite system or constellation: the Ground Sampling Distance (GSD), Modulation Transfer Function (MTF), and Signal to Noise (SNR). Combined with appropriate post-processing, these metrics inform the quality in terms of the ability to process and interpret the image captured for a given application. The National Image Interpretability Rating Scale (NIIRS) provides a subjective rating of an image interpretability [3]. In typical high performance systems, GSD is the most significant driver, contributing > 70% in terms of overall interpretability. A number of quantitative estimations of performance using the NIIRS as a reference have been made, resulting in Table 1.

Table 1: Estimation of NIIRS level using the General Image Quality Equation.

| System | SNR | RER | GSD | NIIRS Estimate |
|---|---|---|---|---|
| Pleiades-A | 22.3 | 0.64 | 0.70 | 5.14 |
| GeoEye-1 | 18.6 | 0.63 | 0.41 | 5.85 |
| WorldView-3 | 25.0 | 0.59 | 0.37 | 5.92 |
| PlanetScope | 30 | 0.55 | 3.7 | 2.67 |
| 3 U modular optics (CCAM) [4] | 20 | 0.55 | 5.0 | 2.25 |
| 3 U deployable optics (HighRes) | 30 | 0.55 | 0.7 | 4.96 |

For the HighRes 3U CubeSat design presented here, the deployable system would enable a rating on the NIIR Scale of 4/5. This would be equivalent to allowing building type, railway traffic and motor vehicle identification [3]. Previously, bus performance limited imaging capability, presenting a barrier to achieving higher resolution, either through absolute pointing or limiting MTF achievable due to jitter. The current generation of CubeSat attitude control systems having demonstrated (15″-42″, 3σ) with low noise [5], now ensure that the image quality is largely driven by the instrument. For Earth monitoring organisations, current developments in deployable systems presents a further boundary pushing capability for CubeSat systems bringing cost-effective platforms increased capacity for high resolution imaging [6, 7, 8]. The development presented in this paper builds on the work previously presented by Schwartz et al. [9].

In this paper, we present an innovative mechanical design for packaging (to fit into the CubeSat volume), deploying and controlling the position of the primary mirror petals. We show the overall performance of focal plane sharpening to improve image quality through extensive numerical simulations under realistic deployment conditions. Finally, we report on the development of a full end-to-end laboratory demonstrator. This demonstration includes an extended scene projection system, an active imaging payload and the mirror co-phasing process using focal plane sharpening.

## 2   PROTOTYPE CUBESAT DESIGN

### 2.1   Optical design

The imaging system the HighRed deployable Cubesat is based around a Cassegrain telescope with a lens assembly to set the correct focal length and to provide aberration correction across the field-of-view (FoV). The longest baseline across our segmented primary mirror is 300 mm, giving diffraction limited resolution of the system is 0.46 arcsec at 550 nm. Typical CMOS imaging detectors have pixels of around 5.5 μm square. To take full advantage of the diffraction limited resolution the focal length of the telescope must be set to match the diffraction limit to 2 pixels on the detector, giving critically (Nyquist) sampled images. These parameters imply a focal length of 4900 mm (focal ratio of f/16.4). At an altitude of 400 km this corresponds to a ground sampling distance of 0.45 m/pixel, and a diffraction limited image resolution of 0.9 m on the ground. A CMOS detector with 2k x 2k pixels will therefore give an instantaneous FoV of $0.9km^2$.

There is a clear trade-off between FoV and sampling. For some applications it may be desirable to accept an undersampled image in order to increase the FoV. Another driver for reducing the sampling would be to increase the signal to noise ratio (SNR) of the images. Achieving a high SNR in diffraction limited images from low earth orbit is extremely challenging due to the very short exposure time needed to avoid blurring of the image by the high speed of the satellite over the ground. While some sort of image stacking is likely to be required to reach a usable SNR, the problem becomes significantly easier if the pixel size on the ground is increased, giving more flux per pixel and allowing longer integration times.

For the purposes of this study we continue to look at a system designed to give a fully sampled image. The telescope must be very compact in order to allow relatively simple deployment from a 3U CubeSat. We have constrained the separation of primary to secondary mirror to a maximum of 200 mm, and the separation from secondary mirror to detector to a maximum of 250 mm. The design shown in Figure 1 produces the required f/16.4 beam at the detector. Different plate scales could be produced by changing the lens elements with a minor re-optimisation of the telescope mirrors. M1 is very close to being a parabola, and consists of the four deployable segments. The primary and secondary mirrors are separated by 200 mm. The lenses – a cemented doublet and a singlet lens – increase the focal length of the telescope while also correcting off-axis aberrations and field curvature across the visible wavelength band. All lens surfaces are spherical, and the lens materials are N-BK7 and N-SF6, for which radiation hardened equivalents are available. This system gives excellent diffraction limited performance across the FoV and across the visible wavelength band. The resulting point spread function (PSF) is shown in Figure 2, illustrating the four lobed structure which is an inevitable consequence of our segmented primary mirror.

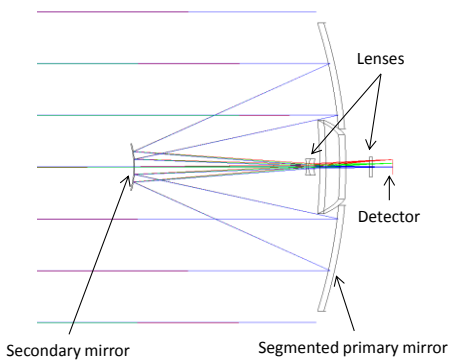
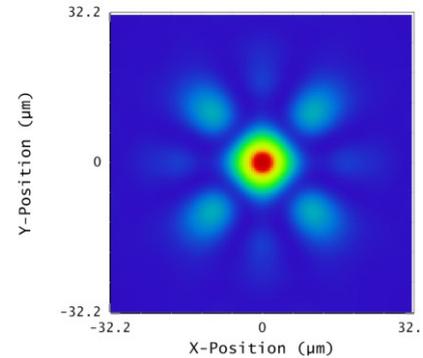

Figure 1: Two different views of the optical design.

Figure 2: The diffraction limited point-spread function.

Analysis shows that the system is extremely sensitive to small misalignments of the segments. This is not surprising when we consider that we are trying to form a single optical surface of diffraction limited performance. The sensitivity to tilt and focus allows us to estimate the resolution of movement required in these axes to achieve co-phasing of the segments. Modelling in Zemax OpticStudio was used to derive the positioning resolution of the three actuators used to manipulate each segment as shown in Table 2. The deployment repeatability figures are set to ensure that when the segments are aligned then folded and re-deployed, they will return to a position sufficiently close to the original aligned position to allow the co-phasing algorithms to work efficiently to fine tune the alignment.

There is a very high sensitivity to decenter of the segments – even a 1 micron decenter has a significant effect. This is a result of the very fast focal ratio of the primary mirror (f/0.75). This misalignment can be compensated to some degree by adjusting the tilt and focus of the segment. As a result, the decenter of the segments needs to be correct to within around 20 micron. This places a tight requirement on the initial alignment of the system and on the repeatability of positioning defined by the hinges of the deployment mechanism. This sensitivity to decenter can be reduced by increasing the radius of curvature of the primary mirror. This would require an increase in the separation between M1 and M2, making the telescope longer. This would require a more complex deployment mechanism for M2, but does have a dramatic effect on the sensitivity, with tolerances relaxed by a factor of ~6.5 if the M1-M2 distance is doubled to 400mm. Such a change would do nothing to relax the resolution requirements for the tilt and focus adjustments of the segments. These sensitivities are independent of the optical design of the telescope, and simply reflect the deviation of the surface of the segment relative to the nominal optical surface.

Table 2: Derived mechanical positioning requirements.

| Orientation | Adjustment resolution | Adjustment stroke | Deployment repeatability |
|---|---|---|---|
| Tip | ± λ/14 (± 45 nm) | 1 mm | ± 10 μm |
| Tilt | ± λ/14 (± 45 nm) | 1 mm | ± 10 μm |
| Piston | ± λ/14 (± 45 nm) | 1 mm | ± 10 μm |

## 2.2 Mechanical design

The key mechanical engineering challenges are: 1) having accurate positioning mechanisms both to deploy and align the mirror segments, 2) having an accurate metrology system that enables the segments to be aligned and co-phased so that they perform as a single monolithic mirror, 3) being able to package the actuator mechanisms and metrology systems in the limited volume.

Figure 3 shows an overview of the prototype CubeSat system and its major sub-components. A modular integration of the CubeSat systems described below was developed for the prototype laboratory demonstrator. This allowed easy access to all the various components for inspection, alignment and modification during the test programme. All structures are predominantly aluminium (including the mirrors) to ensure any differential thermal contraction effects are minimised. Special features and shims for component alignment and adjustment were incorporated into the mirrors and sensors. Much of the integration work was the adjustment of machinable shims to set the baseline mirror positions. The secondary mirror was supported above the CubeSat main structure on an optical bench. The deployment and automatic adjustment of this mirror is beyond the scope of the current work.

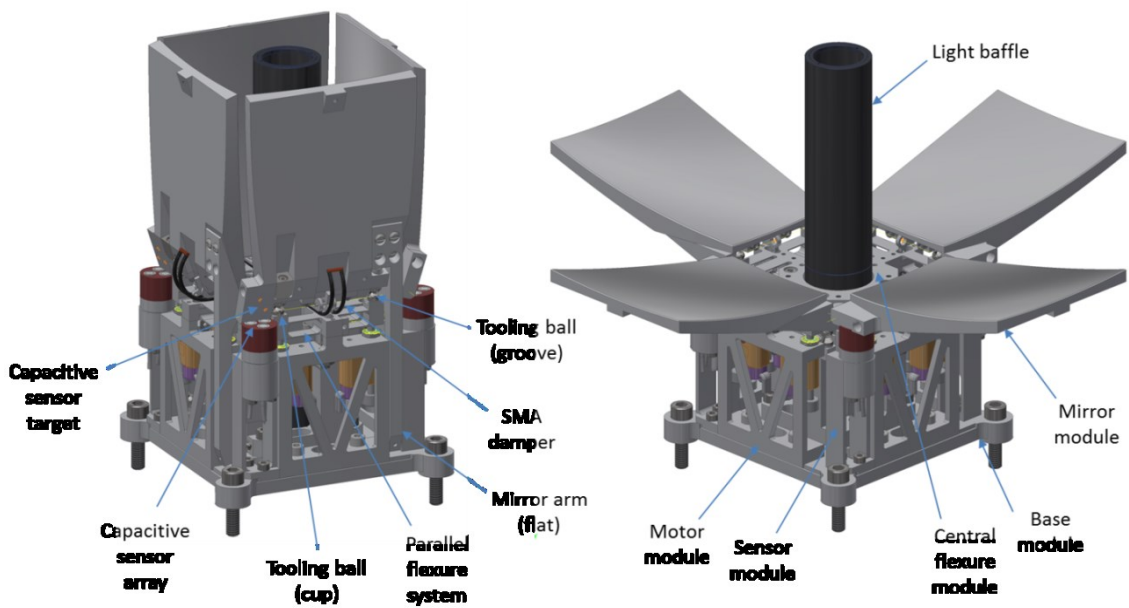

Figure 3: Overview of the mechanical design of the prototype CubeSat.

The mirror positioning mechanisms on this prototype use a floating hinge design that was previously developed at UKATC [9]. Each mirror is mounted on a set of three parallel flexures that are in contact with 3 piezo-driven lead screw motors. Figure 4 shows where the motors contact the mirror petal. The 3 motors enable each mirror to be manipulated in tip, tilt and piston. The motor chosen for this application is the Newport 8354 Tiny Picomotor. They have a 30 nm positioning resolution which is sufficient to meet the specification. Other favourable attributes of these motors are their high actuation force (13 N) and their ability to hold position while powered off.

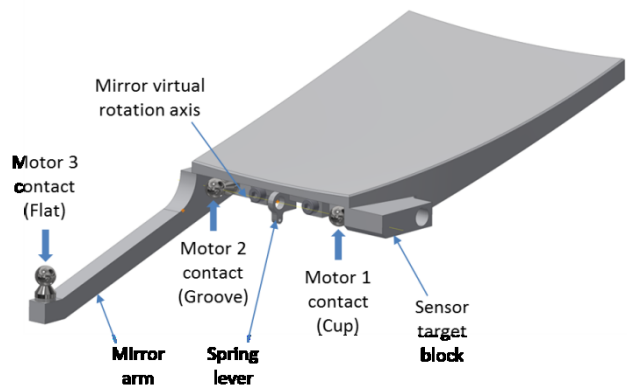

Figure 4: Mirror petal contact points.

Each mirror is kinematically mounted onto the flexures using thee spherical contacts in a cup, groove, flat configuration. These kinematic features are incorporated into the flexure system (not shown). Threaded tooling balls were used for the spherical contact because they have a very low surface roughness and excellent surface form. The tooling balls are made of Stainless Steel AISI 440C. This material is ferro-magnetic and therefore allows the balls to be held in the kinematic mounts using small permanent magnets. These magnets ensure a positive and repeatable contact for each mirror after deployment. The cup and groove mounts form a virtual axis of rotation for each mirror that defines its movement from a stowed to a deployed configuration. When deployed, an 'arm' on each mirror creates the third contact. The final satellite would require the mirrors to be clamped in the stowed position to resist launch loads; this was not considered in the present study.

The deployment motion is provided by an extension spring. It is mounted on a short lever arm that is designed to produce zero torque on the deployment axis when the mirror is in its final position. This is important so that the spring does not interfere with the tip-tilt adjustments on the mirror.

Spring deployment mechanisms can cause undesirable shock or vibration issues when released. These undamped motions can cause repeatability issues in precision mechanisms. To prevent this from happening the mirror motion is controlled by the use of a Shape Memory Alloy (SMA), in this case Nitinol wire. The SMA wire has the ability to return to a predetermined shape on heating to modest temperatures, typically 60-80 °C. The wire is deformed at room temperature to restrain the mirror in its stowed position. When a current is passed through the wire, Ohmic heating causes the wire to straighten in a slow and controlled manner. This motion is used to oppose the spring force and gently settle the mirrors into their kinematic mounts. This deployment method was developed at UKATC [10] as a novel and repeatable damping mechanism for a moveable joint. The wire is positioned so that when fully straight it is no longer in contact with the mirror substrate; this ensures it does not interfere with the mirror manipulations during alignment.

Feedback on the position of each mirror is provided by MicroEpsilon CapaNCDT CSE05 capacitive sensors. These sensors provide static displacement measurements with sub-nm resolution over a measurement range of 0.5 mm. The sensors are mounted in a cluster of three next to each mirror, allowing all three degrees of freedom (tip, tilt and piston) to be measured for each mirror. The sensors directly target the back face of the mirror substrate. This sensor target surface formed the reference plane for the diamond machining of the mirror face and so is has a known relation to the optical path. Furthermore, the surface was diamond machined and has a comparable surface finish to the mirror, making it an excellent choice for a capacitive sensor target.

## 3  CO-PHASING METHODS

### 3.1  Modelling

In order to validate the proposed design and control strategy, a full end-to-end model was developed. This numerical simulation is based on OOMAO, a Matlab toolbox dedicated to Adaptive Optics (AO) systems [11, 12]. OOMAO is based on a small set of classes representing the light source, telescope, wavefront sensor, deformable mirror and an imager of an active or adaptive optics system. After the initial deployment of the 4 mirror segments, and given their associated positioning error, the Point-Spread Function (PSF) will be highly aberrated (see Figure 5).

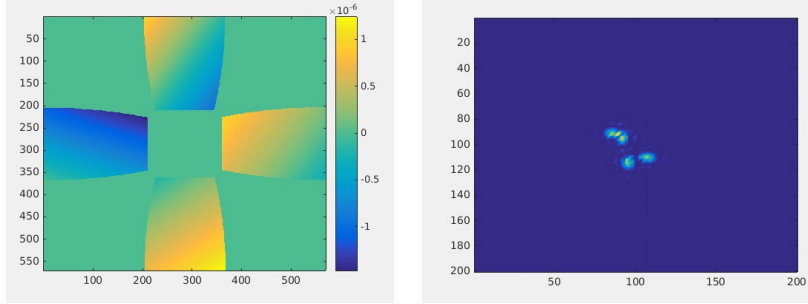

Figure 5: Illustration of initial wavefront aberration (left); the associated PSF (right).

The PSF is the Fourier transform (*TF*) of the input electromagnetic wave: $PSF = |TF[Pe^{i\varphi}]|^2$, where *P* is the pupil function and *φ* the phase. In our study, the wavefront aberrations *φ* are created by the petal misalignments in piston, tip and tilt. For extended objects (e.g. in an EO scenario) and for systems with field-invariant PSFs, the image *I* is simply the convolution (noted *) between the object (or scene) *o* and the PSF p: I(x,y)=o*p. The validity of the field-invariant PSF was verified using OpticStudio simulations, and this assumption will be therefore used in our simulation results.

For our simulations, we used a total of 17 high-resolution EO (see Figure 6) scenes taken from a plane. They contain mostly farmland and urban areas with varied levels of cloud coverage, contrast and brightness levels. The contrast, defined as $C_M = (max_I - min_I)/(max_I + min_I)$ of the images *I*, varies between 0.2 for a scene containing low level clouds to 0.85 for an urban area.

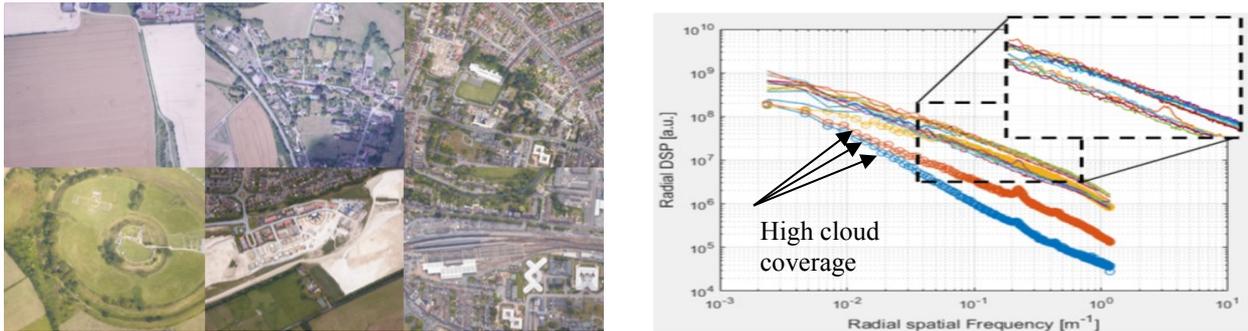

Figure 6: Illustration of high-resolution images (left) and their associated power spectrum (right).

### 3.2 Focal plane sharpening

The image formation process does not retain unambiguously both phase and amplitude of the incoming wave. A common approach to aberration correction involves indirect measurement, whereby wavefront phase is not measured directly with a dedicated sensor, but rather via a sequence of images. The sequence of images is obtained by applying a different and pre-determined combination of modes to the active elements. For each of these images, an image quality metric, such as total intensity or image sharpness, is calculated and used in an optimisation process to find the optimum correction. This method, known as Focal Plane Sharpening (FPS), involves simpler hardware implementations than direct wavefront sensing systems. The indirect measurement process is by nature iterative and is typically applied in fields such as microscopy or laser communications. It is generally not applicable for applications where the aberrations change rapidly (e.g. astronomy). A number of image quality metrics were studied, however for brevity results presented here are shown using ensquared energy: $F_{sq} = \sum_x \sum_y [I(x,y)]^2 / (\sum_x \sum_y I(x,y))^2$.

For this phase of the project we are concentrating on static corrections only. There are no limitations on the number of iterations used to perform the optimisation. We use a modified version of the built-in Matlab function (used to find minimum of unconstrained multivariable function using derivative-free method) to add boundaries. These boundaries restrict the amount of Piston, Tip and Tilt (PTT) to realistic values, such that it searches for the minimum of a problem specified by: $\min_{x_{min} \leq x \leq x_{max}} f(x)$, where *f(x)* is a function (i.e. image metric) that returns a scalar and *x* is a vector containing the PTT values of each mirror segments.

Ideally, a metric should have the following characteristics: (1) ability to measure large aberrations; (2) ability to measure small aberrations; (3) monotonic decrease, (4) min value for no aberrations; (5) insensitive to scene (i.e. metric only sensitive to wavefront); (6) insensitive to noise and background signal; (7) not computational intensive. Finding a metric with all of these characteristics is not realistic and using FPS will come with some limitations. Figure 7 and Figure 8 illustrated how capture range (defined by the full-width half maximum of the metric response curve) and noise sensitivity respectively can be tuned by selecting the metric. It is important to note that as a general rule of thumb, increasing the capture range of the metric will increase its sensitivity to noise. Metrics 1 to 3 represent variations over ensquared energy. Metrics 4 to 9 represent variations of the spatial frequency filtering. Finally, metric 10 represent the image variance and 11 based on a Haar wavelet metric.

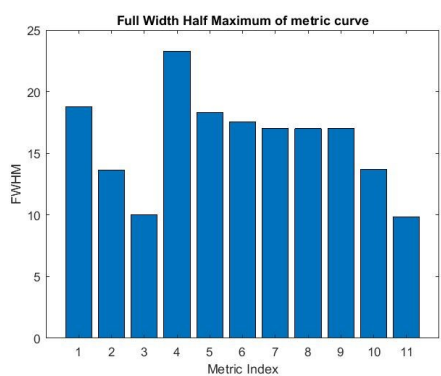
Figure 7: The full width half maximum of different metrics.

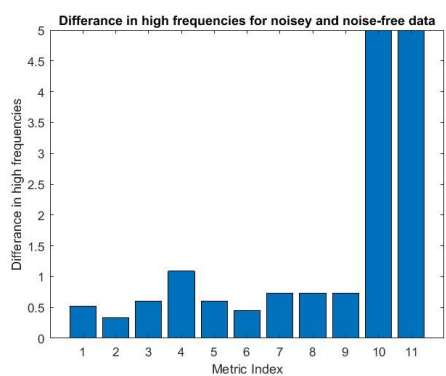
Figure 8: Sensitivity of metrics to noise [arbitrary units].

On-board intelligence and computing is a developing field for CubeSats, driven by the high data rate outputs of next generation sensors and limited bandwidths/power of the system as a whole. With current technology, we have found that there are no major showstoppers in the implementation of FPS algorithms on a CubeSat. However, complex metrics - such as Fourier filtering of complexity $4\log(n)n^2$ - will undoubtedly be more challenging to implement and may limit the lifetime of the satellite by requiring a substantial amount of energy to compute.

### 3.3  Co-phasing simulation results

To create initial aberration, random petal deployment aberrations are generated (see Figure 5). This wavefront is then used as an initial aberration and corrected by moving the mirror segments. The pixel size of the detector assumes 1 or 2 pixel in λ/D (both cases were studied), where D is the diameter of the aperture and λ = 550 nm (i.e. monochromatic simulation). The initial deployment error is set by a uniform distribution between ±2.5 μm PV for tip-tilt (PV given by the edge-to-edge values of 1 petal) and ±1 μm PV for piston; consistent with our previous laboratory tests [9]. Piston values are smaller than tip-tilt because all 3 positions need to be misaligned by the same amount to create a global piston on a mirror segment.

**Point-source**. In this paragraph we study the correction quality of FPS on point-source. Although the CubeSat is ultimately meant for EO, the calibration process (i.e. just after launch) can more easily done on a fixed and know object such as a bright star or the limb of the Moon. We performed simulations of over 500 segment deployments. The average initial Strehl ratio (SR) is approximately 20%; this error is sufficient to have the 4 PSFs created by each individual mirror segments to appear separated on the detector. In high-flux conditions, when the star used is very bright and the electronic noise is low, the average SR after optimisation is 90%. Figure 9 shows a typical example of the image quality before and after correction: the improvement in the PSF quality can be clearly seen. Using smaller pixels to critically sample the PSF (i.e. 2 pixels per λ/D instead of 1) does not noticeably change the final results.

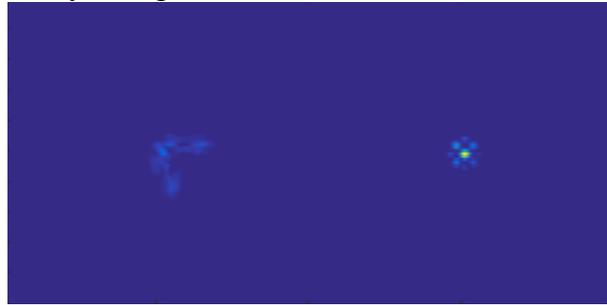

Figure 9: Example of after initial deployment (left) and after phasing (right).

**Extended object**. We also studied the correction quality of FPS on an extended scene, where the PSF is convolved with the object. As with a point-source, we are able to obtain diffraction limited performance for most cases. Figure 10 show a typical example of the diffracted performance reached by the FPS algorithm. However, for images with very poor contrast the final correction quality is also very poor. As a matter of fact, we have found a very strong correlation (>70%) between the final SR and the object contrast ratio. We have shown that a contrast >0.4 should be sufficient for the algorithm to reach diffraction limited performance.

The read-out noise (RON) is the amount of noise generated by electronics as the charge present in the pixels is transferred to the camera. The noise is added every time an image is taken; it will be different for every image taken during the optimisation process. However, since we are using a metric on the entire image (e.g. ensquared energy or Fourier filtering), we have found that RON has very limited impact on the final performance if RON<50$e^-$.

The signal detected by the system includes the light reflected from the ground and light scattered by the atmosphere, particularly at shorter wavelengths. To a first order, the scattered light by the atmosphere will add a constant value to the detected light throughout the image. We have found that, apart from reducing the contrast of the image, scattered light only marginally affect the performance of the FPS algorithm.

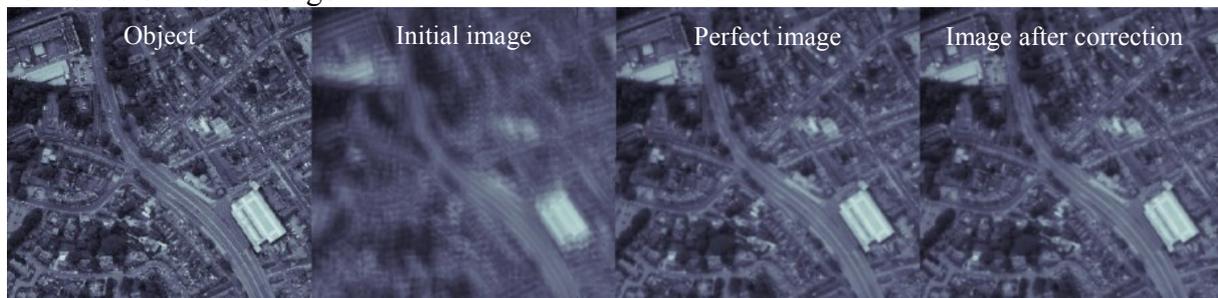

Figure 10: Example of correction quality on extended object.

**Maintaining alignment during operation**. A concept for maintaining the alignment during operations (i.e. with a moving ground scene) was proposed and validated through numerical simulations. The algorithm consists in the following steps: (1) Take image with the unknown mirror misalignments and record metric value Mz; (2) Apply an aberration of a known mode (e.g. piston) and of known amplitude b and record the metric value M+; (3) Apply the same aberration but with opposite amplitude –b and record the metric value M-; (4) Calculate the correction required using a quadratic fit $a = -\frac{b(M_+ - M_-)}{2M_+ - 4M_z + 2M_-}$; (5) Apply the calculated correction $a$ to the mirror mode; (6) Repeat for all modes. This algorithm gives excellent results (i.e. diffraction limited Strehl ratios) for small aberrations. Addition refinements will be required if the drift in PTT of the segments during operation (i.e. after calibration and during EO) are large in the real system. We have found that if the Strehl Ratio is higher than 50-60%, FPS can still achieve diffraction limited performance.

### 3.4 Deconvolution

The shape of the pupil (4 separated segments) leads to a PSF with high side lobes (approx. 25% of central lobe). This clearly impacts the final image where 'waffle-like' features can be identified. This in turn will impact the interpretability of the acquired images (NIIRS). We show that deconvolution can be used to improve the CubeSat image capacity and that there is not showstopper created by the unique shape of the pupil (e.g. by creating unseen spatial frequencies). More details can be found in [13, 14]. Figure 11 shows an illustration of: (1) the initial object/scene, (2) what impact an optical system without any wavefront aberrations has on the image and (3) the improvements that can be obtained by Wiener reconstruction. The Wiener estimation leads to nearly identical image quality both for the images captures using a circular aperture and using the 4-segment aperture.

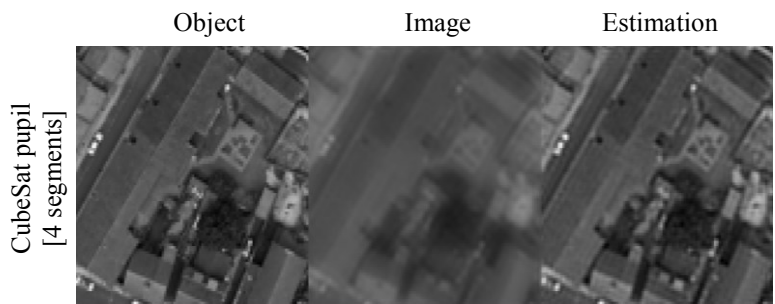

Figure 11: Illustration of the object, image and improvements obtained by Wiener estimation.

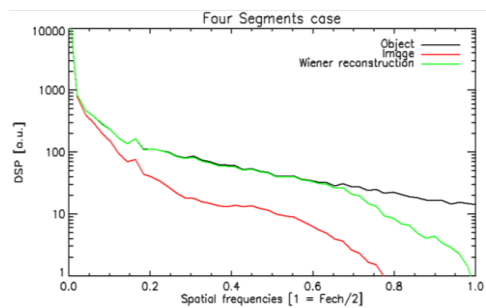

Figure 12: Power Spectral Density for the CubeSat aperture.

Figure 12 more clearly describes the impact of the pupil shape and the performance of the algorithm. It shows the Power Spectral Density (PSD) the CubeSat aperture. It also depicts the PSD for the initial object (black), the image (red) and reconstructed image (green). The only difference can be seen on the high spatial frequencies of the reconstructed images. In low-level noise, we observe good performance with most of the high spatial frequencies retrieved. In addition, the performance between a full circular aperture used as a point of reference (data not shown) and the actual 4-segment aperture is very comparable. Although Wiener filtering attempts to minimize the impact of noise at frequencies which have a poor signal-to-noise ratio, noise will still have considerable impact on the deconvolved images. In the presence of medium noise levels, the performance is naturally degraded and would require additional tuning to improve results. However, this clearly demonstrates that there are no showstoppers in applying deconvolution to image acquired using a diluted pupil such as the one we are using.

## 4 LABORATORY DEMONSTRATION

### 4.1 CubeSat test rig design

The purpose of the demonstrator is to test the co-phasing of a representative segmented mirror. This includes demonstration of the deployment of the segments from a stowed position and

*The 4S Symposium 2018 – N. Schwartz* 9

measurements of the repeatability of the segment positioning after deployment. The segments will be actuated to allow movement with sufficient resolution and range to allow the mirrors to be co-phased. A complete imaging system was constructed, along with test equipment capable of projecting source images into the system. This is to allow testing of co-phasing of the system to be carried out with the mirrors controlled by the algorithms described above.

Deployment and/or actuation of the secondary mirror are not within the scope of this phase of the project. Similarly, the design of parts of the system other than the primary mirror segment deployment and actuation assembly are not required to be fully representative of the implementation that would be required to fit within a CubeSat.

The system is designed to provide the same sampling relative to the diffraction limit as would be required in the real system. The residual wavefront error once the segments are co-phased must be $\leq \lambda/14$ to provide diffraction limited imaging. Assuming that we wish to fully sample the diffraction limit of the system then we need 2 pixels on the detector to map to an angular resolution of $\lambda/D$. Taking $D$ as 300 mm and $\lambda$ as 550 nm we need a pixel scale on the detector of 0.91 µrad/pixel. We selected a Point Grey Flea3 camera for these experiments, which uses an E2V high sensitivity CMOS detector with 5.3 µm pixels. This requires a focal length of 5.8 m, or a focal ratio of f/19.4 (focal ratio based on 300 mm circular aperture).

Since we are not limited by the space constraints of a real Cubesat system we can achieve this using a simple Cassegrain telescope design with no additional correcting lenses required. The primary mirror was constrained to be a parabola to allow interferometric tests of these mirrors alone. The focal plane can be produced 250 mm behind the surface of the primary mirror, allowing the camera to be mounted behind all the deployment and mirror segment alignment mechanisms. This back focal distance also allows for the inclusion of a beamsplitter, allowing the light to simultaneously be sent to a wavefront sensor as shown in Figure 13. This provides the capability for independent optical metrology of the segment positions.

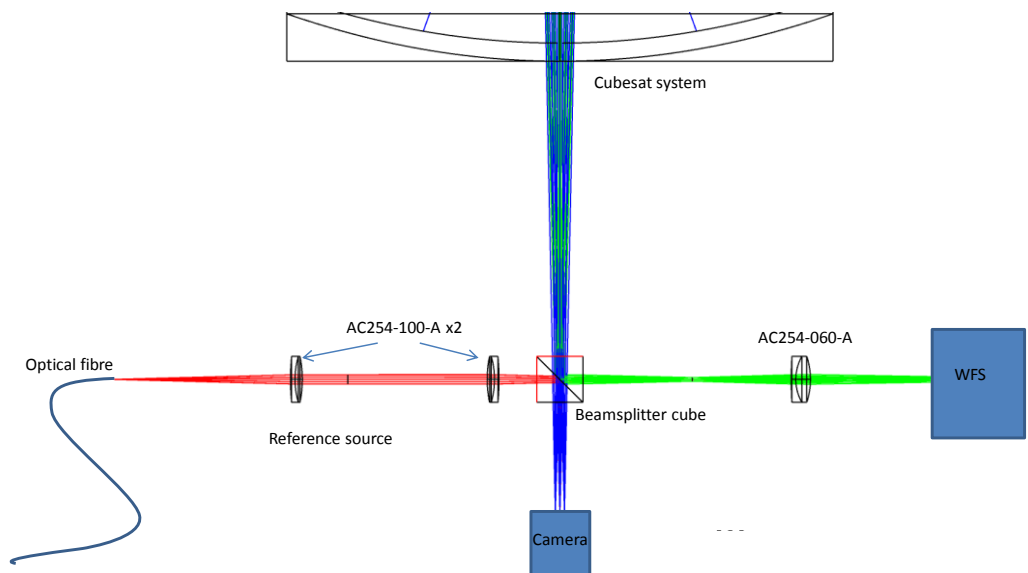

Figure 13: The camera and wavefront sensor assembly in the test bench

The demonstrator (see Figure 14) is mounted in an upwards looking configuration so that all mirror petals experienced the same gravitational forces during deployment and adjustment. An Orion Optics VX12L Newtonian telescope is mounted above the demonstrator to act as a 300 mm collimator as shown in Figure 14. The telescope was supplied with a metrology report giving the reflected wavefront from the primary mirror as 73.3 nm P-V and 7.6 nm RMS.

The input target images are displayed on a Forth Dimension Displays reflective LCD QXGA-R9 with a polarising beamsplitter and LED used to configure this as an emissive display system. This display has 2048x1536 pixels with a pixel size of 8.2 micron. To ensure that the pixellation of the target image has no significant impact on the output image, we need to match 3 pixels of the display to $\lambda/D$. This means that the display must be located at an f/45 focal plane. We therefore need additional optics to couple this to the telescope system. This is achieved using a single negative achromatic doublet combined with a plano-convex lens. Over the FoV that we require this system gives wavefront errors of less than $\lambda/14$, and is diffraction limited provided that the bandwidth of the illuminating light source is kept to no more than about 50nm.

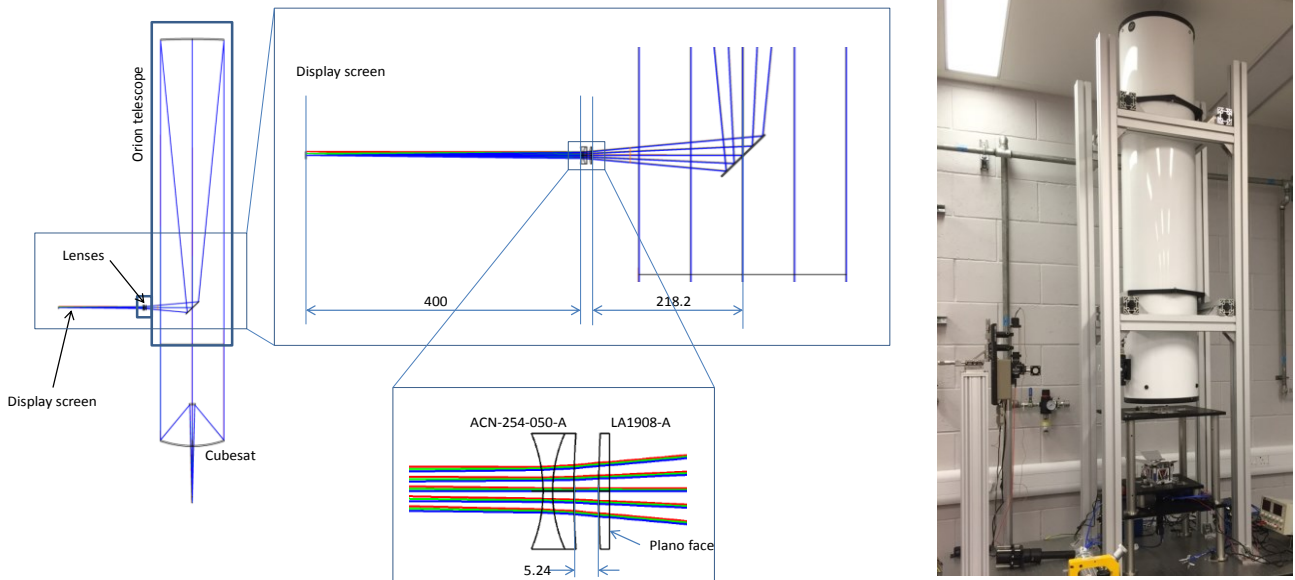

Figure 14: The optical layout of the test bench (left). Laboratory demonstrator (right).

## 4.2 System optical alignment

The mirrors for the Cubesat assembly were manufactured from aluminium with optical surfaces formed by diamond turning. There were some issues during manufacture which resulted in some surface form errors at relatively high spatial frequencies. In order to assess the impact of these errors a single segment was set up on the bench with the secondary mirror and a camera to form an off-axis imaging system. The 100 mm diameter collimated beam from an interferometer was used as an input light source. Initially a return sphere was included in the beam in place of the secondary mirror, providing a double pass measurement of the form error on the segment. The angle of the parabolic segment was adjusted relative to the collimated beam to minimise the astigmatism. The optical axis of the segment determined in this way was found to be misaligned to the mechanical axis of the segment by 2.4 mrad, which would subsequently need to be compensated during the alignment process to the test bench chassis. With the secondary mirror in place a camera was placed at the resulting focal plane to examine the point source image. This image does show a bright diffraction limited core, but this is surrounded by a halo of speckles as shown in Figure 15. The plots in Figure 16 show the circularly averaged intensity profile and the encircled energy profile of this spot image. The camera was replaced with a Shack Hartmann wavefront sensor with a collimating lens. The measured wavefront from this was included in a Zemax raytracing model of the test setup to generate the modelled encircled energy profile included in Figure 16. The spot images generated by this optical model also showed a speckle pattern which appeared qualitatively similar to that observed in our measurements.

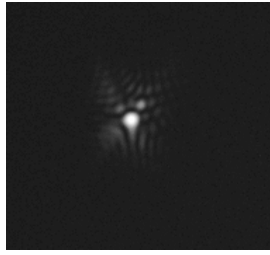

Figure 15: The spot image seen from a single segment with the secondary mirror.

Three of the mirror segments show broadly similar performance levels. The fourth segment has a much larger astigmatism error, believed to be caused by over-constraint of the substrate during diamond machining, and is not considered to be usable for co-phasing tests. The tests that can be carried out using the current set of mirror segments is somewhat reduced from the original aims of the project. The diffraction limited core in the point source images will allow co-phasing of up to three mirrors using a point source image. The large proportion of flux in the speckles around this core will significantly reduce the contrast in images of extended scenes, making co-phasing based on extended scenes significantly more challenging than expected.

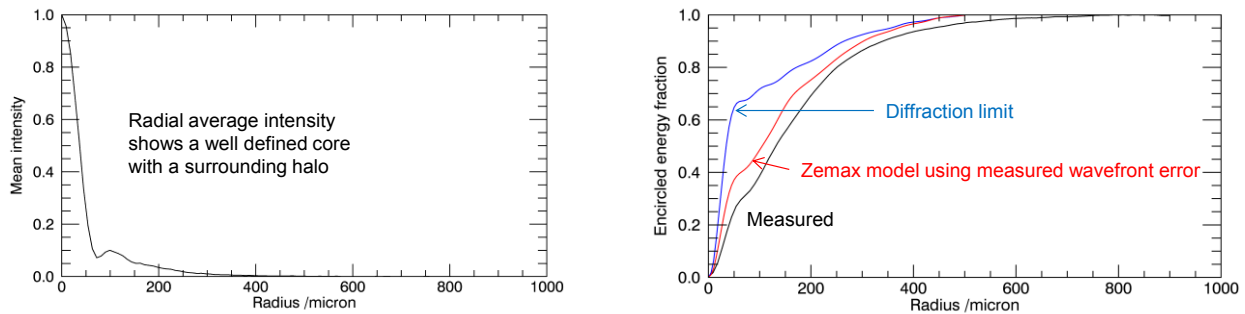

Figure 16: (left) The circularly averaged intensity profile of the spot image shown in Figure 15. (right) The encircled energy profile of this spot as measured (black), as modelled in Zemax using the measured wavefront error (red) and in the diffraction limit.

The segments were aligned relative to the Cubesat chassis and secondary mirror by using shims to adjust their lateral positions. For each segment an iterative process was carried out to optimise the spot images seen on the camera, with the segment tip-tilt and focus positions adjusted for each shim using the three actuators. The adjustments that were required were larger than initially expected due to the misalignment described above between the optical and mechanical axes of the segments. Adjustments of up to 600 micron were required, and were carried out to an accuracy of around 40 micron. Once the three usable segments were aligned in this way, the three resulting spot images were brought into alignment with one another on the camera, defining the start point for the co-phasing experiments.

## 4.3 Deployment repeatability

For the co-phasing algorithms to work it is required that the mirrors are deployed into their start position with adequate repeatability. As described in Section 2.1, the specification for this repeatability has been set at ±10 μm. To test the deployment performance the mirrors were retracted and deployed into their kinematic mounts 500 times. Each time the mirror settled into its mount a reading from each of the three capacitive sensors was taken. Figure 17 shows the distribution of the measurement data for each sensor. The red line shows a normal distribution fit to the data with mean subtracted. From the figures it can be seen that in all instances the 2σ variation (95% confidence) around the mean is well within the required repeatability tolerance of ±10 μm. Sensors 1 and 2 show a close approximation to a normal distribution, but sensor 3 has a bias toward the left tail. The reason for this bias is not clear and more testing is required to determine its source.

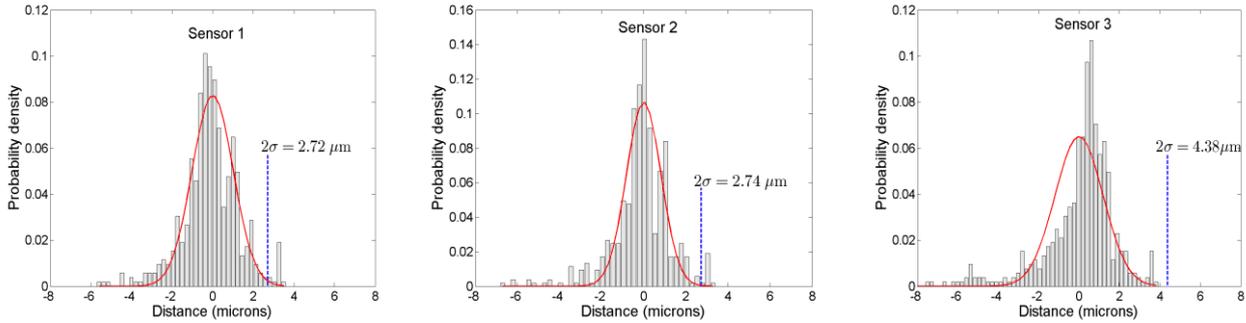

Figure 17: Deployment repeatability histograms for the three sensors. A best fit normal distribution for each chart is shown in red.

### 4.4 Performance of the motors and position sensors

The resolution of the mirror manipulation is a key mechanical requirement to enable the mirror co-phasing. Table 2 shows that the resolution of the mirror manipulation is required to be better than 45 nm at the three motor locations. Figure 18 (left) shows the mirror movement at each of the three sensor locations for a 1000 step movement of motor 1. This sensor reading translates into approximately 7.3 µm displacement at the motor contact point. The theoretical single-step resolution is therefore well below the 45 nm requirement. However in practice the motors display hysteresis, backlash and a load/displacement dependency, which mean that it is not practical to use single-step commands as part of the mirror control, but the benefit of this very high resolution can be utilised over larger motor movements.

Figure 18 (right) shows the first 0.4 seconds of data from the motor 1 test. This shows that the recorded peak-to-valley noise floor of the sensors at 100 Hz is less than 2 nm. This is considered adequately low to enable to perform the co-phasing work described in Section 3.

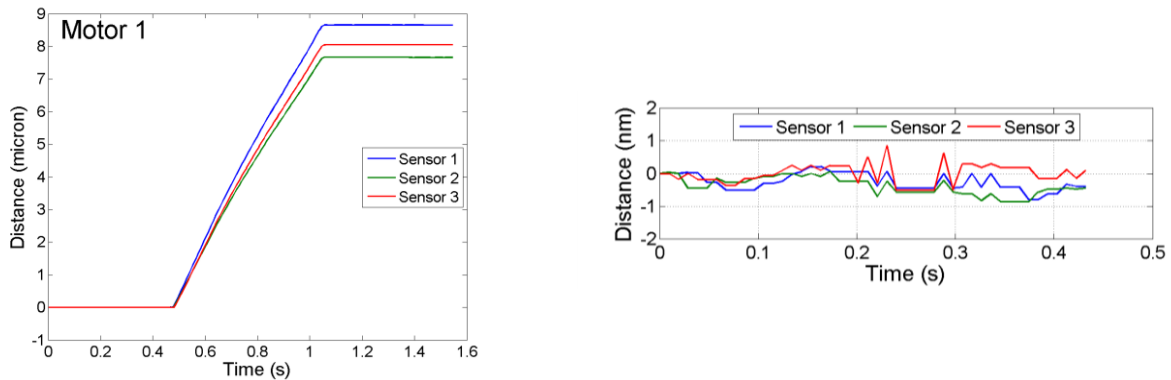

Figure 18: Sensor measurement data for 1000 step movement on motors 1 (left). Other motors have similar behaviours. Noise floor data from the motor 1 test prior to motor actuation (right).

### 4.5 Closed-loop active optics

In this section, we use the capacitive sensor and the PicoMotor in conjunction to drive the mirrors into a desired position. We use a classical integrator control: $u_{n+1} = u_n - gM_{control}s_k$, where $M_{control}$ is the control matrix, $g$ is the uniform scalar gain of the integrator and is equal to 0.5, $u_n$ is the PicoMotors control vector at iteration step $n$, and $s_k$ is the capacitive sensor measurements. $M_{control}$ is calculated from the generalised pseudo-inverse of the interaction matrix which itself is measured by recording the response on the sensors of each motor sequentially. Figure 19 shows the 3 capacitive channel measurements taken from mirror segment number 1. The initial error to the desired position is between 5 µm to 12 µm. Despite the hysteresis and backlash in the motor we are able to converge to a final error < 25 nm.

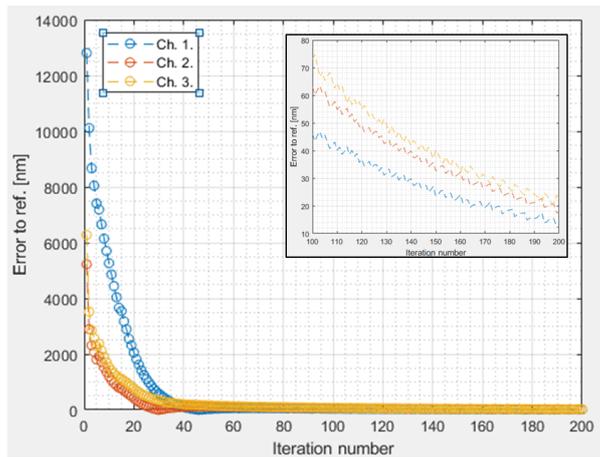

Figure 19: Sensor measurements as a function of iteration step. Inset shows a zoom between iterations 100 and 200.

## 5   CONCLUSIONS AND NEXT STEPS

A concept for HighRes, a high-resolution Earth Observation CubeSat using a deployable primary mirror was presented. A laboratory opto-mechanical prototype of this concept was developed and discussed in this paper. We demonstrated the feasibility of having a segmented primary mirror that can be deployed; its position measured and manipulated with the level of accuracy required for the co-phasing. The prototype is packaged within 1.5U CubeSat volume and uses space-compatible hardware. We use a focal plane sharpening algorithm to co-phase the different segments. We demonstrated using end-to-end numerical simulations that it can be a viable method of co-phasing the four mirror segments, both for the initial alignment procedure after launch or during operations with a moving extended scene. Performance of the demonstrator is currently limited by the quality of the mirrors. In order to do a full experimental validation of the co-phasing process we need to improve their quality. We demonstrated that the position of the mirrors, using the capacitive sensors and the PicoMotors in conjunction, to within 25 nm of their optimal position; an accuracy sufficient to achieve diffraction limited imagery.

The next steps in this project are to upgrade the demonstrator to verify the performance of the focal plane sharpness algorithm under realistic experimental conditions. We also plan on improving the algorithm to further explore the possibility of co-phasing a deployable CubeSat using a moving ground scene and when aberrations are more severe. In the longer term, we plan on investigating opto-mechanical designs to deploy the secondary mirror, miniaturise the electronics and especially the drive electronics. Working within a 3U volume severely constrains the design and so we will also investigate the benefit in terms of costs and design of moving to a larger 6U volume or other small satellite platforms.

## 6   ACKNOWLEDGMENTS

The funding of this work by DSTL and UKSA is gratefully acknowledged.